# Purcell enhancement of silicon W centers in circular Bragg grating cavities

Baptiste Lefaucher[1]*, Jean-Baptiste Jager[1], Vincent Calvo[1], Félix Cache[2], Alrik Durand[2], Vincent Jacques[2], Isabelle Robert-Philip[2], Guillaume Cassabois[2], Yoann Baron[3], Frédéric Mazen[3], Sébastien Kerdilès[3], Shay Reboh[3], Anaïs Dréau[2] and Jean-Michel Gérard[1]*

[1]Université Grenoble Alpes, CEA, Grenoble INP, IRIG, PHELIQS, Grenoble, France

[2]Laboratoire Charles Coulomb, Université de Montpellier and CNRS, Montpellier, France

[3]Université Grenoble Alpes, CEA-LETI, Minatec Campus, Grenoble, France

* baptiste.lefaucher@cea.fr, jean-michel.gerard@cea.fr

**ABSTRACT**

Generating single photons on demand in silicon is a challenge to the scalability of silicon-on-insulator integrated quantum photonic chips. While several defects acting as artificial atoms have recently demonstrated an ability to generate antibunched single photons, practical applications require tailoring of their emission through quantum cavity effects. In this work, we perform cavity quantum electrodynamics experiments with ensembles of artificial atoms embedded in silicon-on-insulator microresonators. The emitters under study, known as W color centers, are silicon tri-interstitial defects created upon self-ion implantation and thermal annealing. The resonators consist of circular Bragg grating cavities, designed for moderate Purcell enhancement ($F_p = 12.5$) and efficient luminescence extraction ($\eta_{coll} = 40\%$ for a numerical aperture of 0.26) for W centers located at the mode antinode. When the resonant frequency mode of the cavity is tuned with the zero-phonon transition of the emitters at 1218 nm, we observe a 20-fold enhancement of the zero-phonon line intensity, together with a two-fold decrease of the total relaxation time in time-resolved photoluminescence experiments. Based on finite-difference time-domain simulations, we propose a detailed theoretical analysis of Purcell enhancement for an ensemble of W centers, considering the overlap between the emitters and the resonant cavity mode. We obtain a good agreement with our experimental results assuming a quantum efficiency of 65 ± 10 % for the emitters in bulk silicon. Therefore, W centers open promising perspectives for the development of on-demand sources of single photons, harnessing cavity quantum electrodynamics in silicon photonic chips.

**INTRODUCTION**

Silicon is a leading material platform for scaling up quantum photonics to an advanced level of complexity. [1,2] Over the past ~15 years, the improving quality and quantity of components in silicon-on insulator (SOI) photonic chips have enabled the generation, coherent control and detection of increasingly large quantum states. [3,4] However, an obstacle to further expansion arises from the difficulty to generate single photons on demand in silicon. Indeed, sources based on four-wave mixing in silicon (the main workhorse to date) produce heralded photons in a non-deterministic way and with a limited efficiency in the 5 to 10% range. [1] To overcome this limitation, research efforts have focused on fiber-coupling or hybrid integration of external single-photon sources (SPS), [5] emitting at telecom wavelengths for low absorption in silicon waveguides, such as III-V semiconductor quantum dots (QD) [6,7] and erbium ions in YSO. [8] In spite of tremendous progress in telecom photon generation in various platforms, [9–13] the scalability of these approaches is challenged by coupling losses at interfaces. [14]

A promising alternative route exploits the integration of isolated atoms into the photonic chip, such as erbium ions [15] or isolated defects acting as artificial atoms. [16–18] By embedding one of such emitters in a SOI cavity, one expects to tailor its spontaneous emission (SE) thanks to cavity quantum electrodynamics (CQED) effects, and to control the single-photon generation process. [19] Concerning color centers in Si, particular attention has been devoted to the G center so far. [20–24] Ensembles of G centers have been integrated in a metasurface, micropillars and Mie resonators to improve light extraction out of silicon. [25–27] Cavity enhancement of the zero-phonon emission has been demonstrated for an ensemble of G centers in a microring cavity, [28] and in a 2D photonic crystal (PhC) cavity for a single G center. [29] Recently, an 8-fold acceleration of the SE decay of a single center in a 2D PhC cavity has been demonstrated. [30] However, the emission spectrum of the G center is dominated to ~85% by phonon side-bands (PSB), which complicates further improvement toward GHz-rate emission of indistinguishable photons needed for several applications. [31] For that purpose, the quest for a better artificial atom is still widely open.

The potential of color centers for CQED can be assessed by studying their recombination dynamics. In bulk material, the decay rate $\Gamma_{bulk}$ of a color center is the sum of its zero-phonon emission rate $\Gamma_{ZPL}$ (ZPL = zero phonon line), phonon-assisted emission rate $\Gamma_{PSB}$ and non-radiative decay rate $\Gamma_{nr}$. In an optical cavity resonant with a given transition, the SE is distributed between the resonant mode and a continuum of « leaky » modes. The ratio between the rate of that transition in the resonant mode and in bulk material is defined as the Purcell factor $F_p$. The ratio between the rate of all the radiative transitions in the leaky modes and in bulk material is equal to a factor $\gamma$. In the case of a ZPL-resonant cavity, the decay rate of the color center becomes

$$\Gamma_{cav} = F_p \Gamma_{ZPL} + \gamma(\Gamma_{ZPL} + \Gamma_{PSB}) + \Gamma_{nr} \quad (1)$$

Note that this equation gives the decay rate in bulk material for $F_p = 0$ and $\gamma = 1$. Deterministic single-mode photon emission is achieved for $F_p \Gamma_{ZPL} \gg \gamma(\Gamma_{ZPL} + \Gamma_{PSB}) + \Gamma_{nr}$. It is apparent that this condition is favored when zero-phonon emission is the dominant recombination process in bulk material. This translates to a Debye-Waller factor $\xi = \Gamma_{ZPL}/(\Gamma_{ZPL} + \Gamma_{PSB})$ and quantum efficiency $\eta_{QE} = (\Gamma_{ZPL} + \Gamma_{PSB})/\Gamma_{bulk}$ close to unity, as is the case for state-of-the-art SPS based on semiconductor QDs. [32] In silicon, the newly isolated W center emerges as a good potential candidate for CQED thanks to its high Debye-Waller factor of nearly 40%. [17,33] Ensembles of W centers have already been integrated in waveguide-coupled microring resonators in order to build a SOI circuit-integrated light source. [34] However, precise information about its quantum efficiency is lacking to date.

To probe further the interest of W centers for CQED experiments, we have embedded ensembles of W centers inside circular Bragg grating (CBG) cavities. CBG cavities, which have been widely used in recent years in combination with various emitters such as color centers in diamond, [35,36] semiconductor QDs, [37–39] and 2D materials, [40,41] are attractive because they provide broadband Purcell enhancement and efficient light extraction and collection thanks to their far-field radiation pattern. Here, Purcell-enhancement of W centers is witnessed in photoluminescence (PL) experiments through the observation of both a large enhancement of the collected PL intensity and an acceleration of the PL decay, when the ZPL is in resonance with the mode of the CBG cavity. On the basis of a detailed modelling taking into account the spatial distribution of the emitters in the cavity, we estimate the quantum efficiency of the present ensemble in bulk Si to be of 65 ± 10 %, which confirms the attractiveness of W centers for integrated quantum photonics.

**RESULTS**

**Sample**

The sample was fabricated in a 25 × 25 mm² blank piece of commercial (001)-oriented 220 nm / 2 μm SOI wafer. The top layer was implanted with 70 keV Si ions at a fluence of $10^{12}$ cm$^{-2}$ (**Figure 1a**), and the sample was thermally annealed at 250 °C for 30 minutes in N$_2$ atmosphere (**Figure 1b**). The cavities were then fabricated using electron-beam lithography and reactive ion etching (**Figure 1c**).

The first two steps aim to create a dense ensemble of optically active W centers. [42,43] Under laser excitation, the centers de-excite by emitting zero-phonon PL, phonon-assisted PL and through non-radiative recombination channels (**Figure 1d**). The resulting spectrum at 10 K, shown in **Figure 1e**, consists of a ZPL at 1218 nm and a PSB at lower energy. The ZPL contains a Debye-Waller fraction $\xi \approx 39\%$ of the total emission spectrum, while the first phonon replica (FPR) in the PSB accounts for a fraction $\xi_{FPR} \approx 26\%$ of the emission.

A scanning electron micrograph of a CBG cavity is shown in **Figure 1f**. The cavities consist of a disk of diameter $D$ surrounded by $N$ grating rings of period $p$ and gap-width $w$. Unlike previous works for which a CBG has been combined with a metallic bottom mirror to ensure upwards emission, [38–40] we exploit here the reflectivity of the interface between the SiO$_2$ buried oxide and the Si substrate ($R \approx 16 \%$). In comparison with metallic mirrors or planar Bragg reflectors, our choice simplifies very significantly the nanofabrication process, as no flip chip or wafer bonding step is involved.

In order to facilitate precise measurements of CQED effects as a function of spectral detuning, the cavities were designed to support a moderately broadband resonant mode ($Q \sim 100 - 200$). Due to the high-index-contrast between silicon and the surrounding materials, four rings are enough to reach such a quality factor. [44] The dimensions were then optimized for maximum Purcell acceleration of a single dipole, on the basis of finite-difference time-domain simulations with a commercial software (RSOFT suite by Synopsys). The dipole is located at the center of the cavity and we chose a dipole oriented along the ⟨111⟩ direction as observed experimentally for isolated W centers. [17] A Purcell enhancement factor of 12.5 with a quality factor $Q \approx 160$ were predicted for $D = 970$ nm, $p = 450$ nm and $w = 120$ nm. **Figure 1g** shows the electric field map for the resonant cavity mode for this set of parameters. The mode volume was calculated to be $V \approx 0.65 \, (\lambda/n)^3$. These simulations highlight the good performances of these cavities, in spite of the implementation of this minimal back-mirror, made of a single SiO$_2$/Si interface.

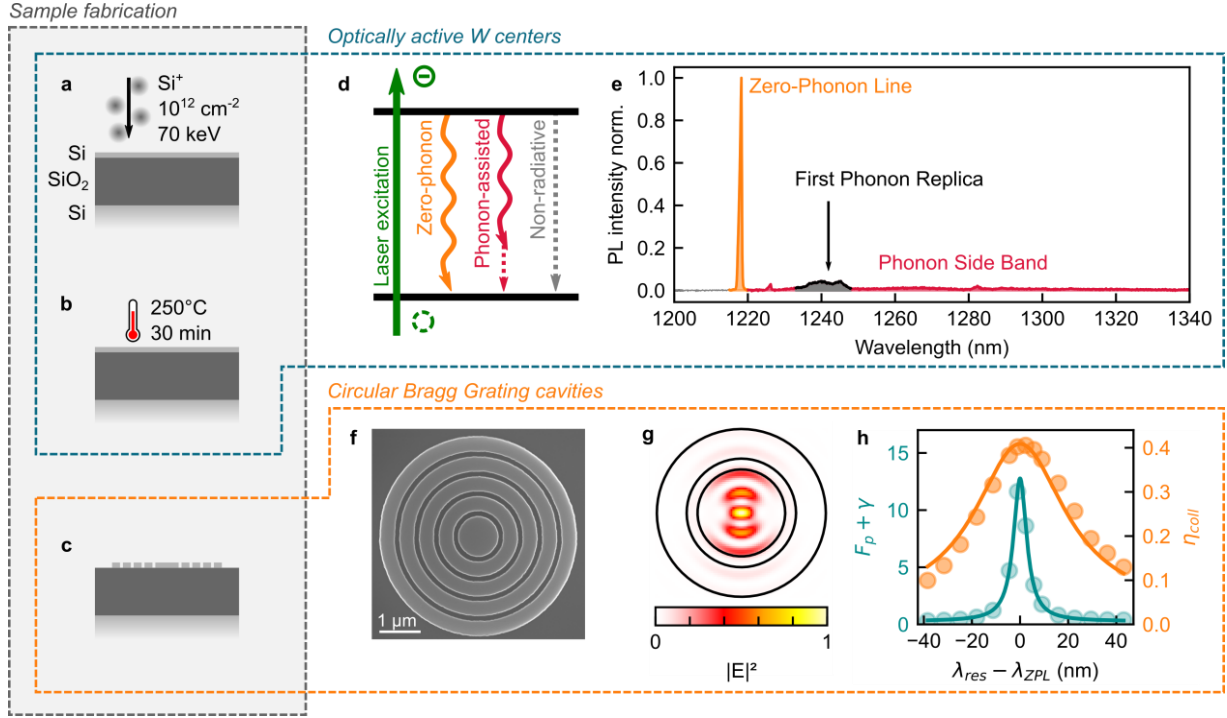

**Figure 1.** Ensemble of W centers in CBG cavities. **a, b, c.** Fabrication of the sample consisting in **a.** silicon ion implantation, **b.** thermal annealing and **c.** nanofabrication using electron-beam lithography and reactive ion etching. **d.** Sketch of W centers recombination channels under above-gap optical excitation. **e.** PL spectrum of an ensemble of W centers in SOI at 10 K. **f.** Top-view scanning electron micrograph of a CBG cavity. **g.** In plane field intensity map of the resonant electromagnetic mode. **h.** Simulated ratio $\Gamma_{cav}/\Gamma_{bulk} = F_p + \gamma$ and collection efficiency $\eta_{coll}$ in $NA = 0.26$ for a single dipole oriented along $\langle 111 \rangle$ at the center of the cavity, as a function of spectral detuning between the resonant mode and the ZPL.

In order to adjust finely the spectral tuning between the resonant mode and the ZPL, an array of cavities was fabricated where the diameter $D$ increases incrementally in 2 nm steps while the other parameters remain constant. The resonant wavelength $\lambda_{res}$ as a function of $D$ was calculated by FDTD. The normalized emission rate at the ZPL wavelength $\Gamma_{cav}/\Gamma_{bulk} = F_p + \gamma$ and collection efficiency $\eta_{coll}$ in a numerical aperture of 0.26 as a function of $D$ were calculated by FDTD as well. The combined results give the emission rate of the dipole (normalized to its value in bulk silicon) and the collection efficiency as a function of detuning, shown in **Figure 1h**. A collection efficiency of 40% is achieved at zero detuning and remains close to this value up to ~10 nm detuning. The solid lines in the figure are fits to Lorentzian functions: the values for large detuning provide estimates for the normalized emission rate into the leaky modes of the structure $\gamma = 0.27 \pm 0.02$ and for the collection efficiency of the leaky modes $\eta_{leaks} = 5 \pm 2\,\%$. The far-field radiation pattern and collection efficiency as a function of numerical aperture are presented in **Figure S1** in the **Supplementary Information**.

**Optical spectroscopy**

Optical spectroscopy measurements at 10 K were performed under 100 MHz pulsed excitation at 485 nm (see the **Methods** section for details). **Figure 2a** shows the PL spectra recorded for cavities providing particular W-mode coupling cases, namely when the cavity is ZPL-resonant (pink spectrum), FPR-resonant (purple) and off-resonant (blue). By « off-resonant », we mean that the resonant mode is far-blueshifted from the ZPL, such that no emitter-mode interaction occurs. A PL spectrum of W in a reference 2D layer is shown for reference (orange). Reflectivity measurements were performed by focusing white light onto the cavities and sending the reflected light to the spectrometer. The reflectivity of the cavities is plotted in the insets to show the spectral position of the resonant mode. When the mode is tuned to the zero-phonon transition wavelength, a 20-fold increase of the ZPL intensity is observed compared to the off-resonant case. In this case, the ZPL contains 90% of the total collected emission. When the mode is tuned to 1240 nm, the whole intensity integrated in the FPR is increased by a factor of 17 and accounts for 79% of the collected emission. This shows the capability of the cavity to enhance broad emission bands. The intensity of the ZPL as a function of detuning with the resonant mode under continuous-wave (CW) excitation is shown in **Figure 2b**. The data are fitted to a Lorentzian profile with a full-width at half-maximum (FWHM) of $8.8 \pm 0.8$ nm. This value is slightly larger than the FWHM of the resonant mode (~7.6 nm), which can be expected when the emitters are pumped below saturation and have a high quantum efficiency.[45] The evolution of PL enhancement as a function of detuning confirms that the exaltation of the ZPL is induced by coupling to the resonant mode. As shown in the **Discussion** section, these observations constitute a first signature of the Purcell effect at play in these cavities.

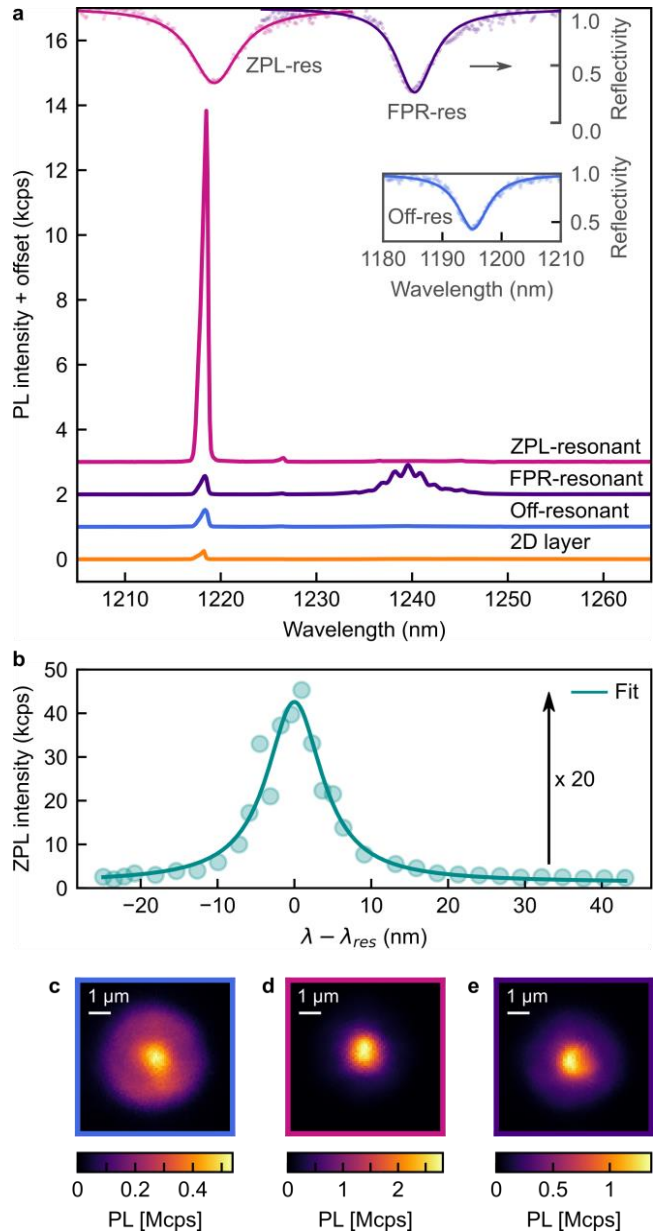

**Figure 2.** Enhancement of the total PL intensity at resonance. **a.** PL spectra of three cavities of diameter 922 nm (blue), 958 nm (pink) and 988 nm (purple), and of the reference 2D layer (orange). White-light reflectivity measurements on top and in the inset, indicate that the cavities are respectively off-resonant (blue reflectivity curve), ZPL-resonant (pink reflectivity curve) and FPR-resonant (purple reflectivity curve). **b.** Intensity of the ZPL as a function of detuning between the resonant mode and the ZPL. **c, d, e.** PL maps of the cavities for an off-resonant cavity (c), a ZPL-resonant cavity (d) and a FPR-resonant cavity (e). Note that (a) and (b) cannot be compared with (c), (d) and (e) in a quantitative manner as the experimental conditions are much different (see the **Methods**).

It is also expected that the emitters-mode coupling modifies the spatial distribution of the emission, as the coupling boosts the emission of the W centers located close to the cavity mode antinode. In order to observe this effect, optical scans of the cavities were performed under CW laser excitation at 532 nm (see the **Methods**). The PL intensity was collected over a broad spectral range using only a 1050 nm longpass filter. **Figure 2c** shows a scan of an off-resonant cavity. The bright spot at the center is attributed to emission from the inner disk of the cavity, while the purple disk is attributed to emission from the surrounding grating rings. **Figure 2d** shows a scan of a ZPL-resonant cavity. The maximum PL intensity is increased by a factor of ~5 compared to the off-resonant case, which is understood as an average increase of the total intensity induced by the exaltation of the ZPL. Moreover, the emission is concentrated close to the center of the cavity, which is expected as the resonant mode is confined in the inner disk. Note that the scans are displayed on different scales for the sake of visibility, but the intensity of the surrounding rings is the same for all the cavities. **Figure 2e** shows a scan of a FPR-resonant cavity, where a ~3-fold increase of the maximum PL intensity is obtained. Achieving enhanced total intensity by enhancement of phonon-assisted PL is made possible by the large spectral width of the resonant mode.

### Time-resolved experiments

Time-resolved PL experiments were performed under 1 MHz pulsed excitation. The PL was collected over the 1200 – 1300 nm spectral range in order to select the W center and reject potential parasitic emitters. **Figure 3a** shows the time traces obtained for the three cavities under study. Qualitatively, the decay of the intensity is the slowest in the off-resonant cavity, faster in the FPR-resonant cavity and the fastest in the ZPL-resonant cavity. The shape of the time-traces is highly multi-exponential, which is expected for an ensemble of emitters. Additionally, the curves feature $\sim\mu s$ time components never reported for W centers. The same curves were obtained when the ZPL was spectrally selected (see **Figure S2** in the **Supplementary Information**). Hence, these slow temporal components cannot be attributed tentatively to parasitic emitters, and must be considered as properties of the present ensemble of W centers.

**Figure 3b** displays the first 80 ns of the time-traces. Over this time window, the decay of the intensity in the off-resonant cavity is nearly mono-exponential, while it is multi-exponential with a steeper initial slope in the ZPL-resonant and FPR-resonant cavities. After ~60 ns, the slope is nearly the same for all the cavities. We attribute this multi-exponential feature to the fact that the emitters are unequally coupled to the resonant mode, leading to a strong dispersion of the radiative emission rate. In spite of this strong dispersion, the data fit reasonably well to a bi-exponential function $I(t) = ae^{-t/\tau_1} + (1-a)e^{-t/\tau_2}$. As a method to characterize the decays, the average time constant $\langle \tau \rangle =$ $a\tau_1 + (1-a)\tau_2$ was calculated for each curve. An average time constant $\langle\tau\rangle = 39 \pm 2$ ns is obtained for the off-resonant cavity, slightly larger than the value of $34 - 35$ ns reported for ensembles in SOI. [17,43] A time $\langle\tau\rangle = 28 \pm 1$ ns is obtained for the FPR-resonant cavity corresponding to an enhancement by a factor $1.4 \pm 0.1$ compared to the off-resonant cavity. A time $\langle\tau\rangle = 19 \pm 1$ ns is obtained for the ZPL-resonant cavity, corresponding to an enhancement by a factor $2 \pm 0.2$. The recombination rate $1/\langle\tau\rangle$ as a function of detuning is shown in **Figure 3c**. The data are fitted to a sum of two Lorentzian functions with an offset. The blue horizontal line in the figure corresponds to the offset, which includes the decay rate of the emitters into the leaky modes and non-radiative recombination channels, $1/\tau_{leaks} + 1/\tau_{nr} = 0.025 \pm 0.001$ ns$^{-1}$. The pink curve is the zero-phonon emission rate into the resonant mode $1/\tau_{ZPL-mode}$, reaching a maximum value of $0.026 \pm 0.002$ ns$^{-1}$ for $\lambda_{res} = 1217.7 \pm 0.2$ nm, with a FWHM of $9.2 \pm 0.9$ nm. The purple curve is the FPR emission rate into the resonant mode $1/\tau_{FPR-mode}$, reaching a maximum value of $0.012 \pm 0.002$ ns$^{-1}$ for $\lambda_{res} = 1240.2 \pm 0.8$ nm, with a FWHM of $14 \pm 4$ nm. These results demonstrate an acceleration of the total decay rate induced by Purcell enhancement of spontaneous emission into the resonant mode.

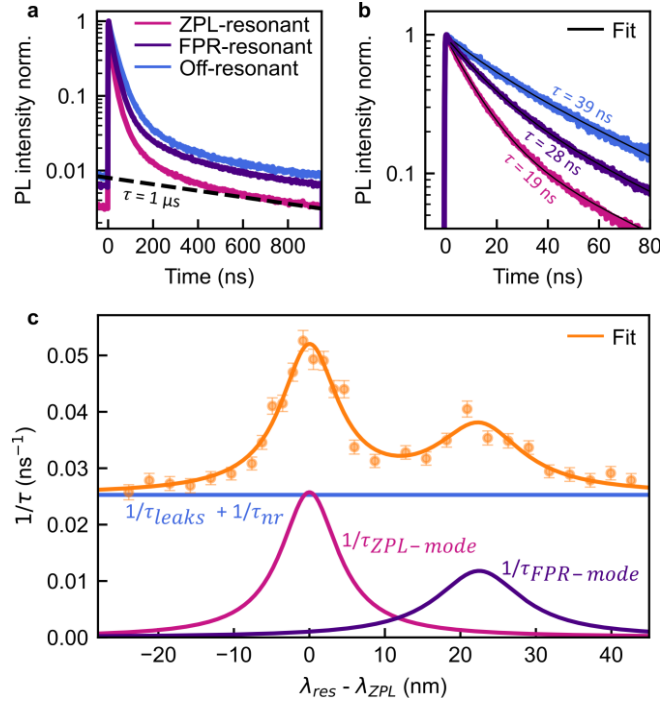

**Figure 3.** PL decay acceleration as observed by time-resolved photoluminescence. **a.** PL decay curves of the ZPL-resonant, FPR-resonant and off-resonant cavities. **b.** Bi-exponential fit of the decay curves over the first 80 ns. **c.** Recombination rate as a function of detuning between the resonant mode and the ZPL. The data are fitted with the sum of two Lorentzian functions with an offset. The total decay rate is interpreted as the sum of the emission rate into the leaky modes $1/\tau_{leaks}$, non-radiative transition rate $1/\tau_{nr}$, ZPL rate into the resonant mode $1/\tau_{ZPL-mode}$ and FPR rate into the resonant mode $1/\tau_{FPR-mode}$.

The experiments were performed with an average power of 0.95 µW (~1.3 W.cm$^{-2}$), far below the saturation threshold of the emitters. Interestingly, the decay rate of the W centers was found to be dependent upon the excitation power. An explanation could be that above-bandgap excitation at high power induces thermal heating, which is known to accelerate the decay of silicon color centers. [46] Nevertheless, the acceleration of the PL decay was clearly observed over the full power range under study, with an acceleration factor almost power-independent. See the **Power study** section in the **Supplementary Information** for more details.

**DISCUSSION**

In this section, we present a theoretical analysis of Purcell enhancement for an ensemble of W centers. The model takes into account the spatial distribution of the W centers and its overlap with the resonant cavity mode. By computing the intensity collected from each emitter, we reconstruct the total intensity and average decay rate as a function of detuning.

The designed CBG cavities have a cylindrical symmetry and support two polarisation-degenerate resonant modes of volume $V \approx 0.65 \, (\lambda/n)^3$ and quality factor $Q \approx 160$. These two figures of merit of the cavity set the maximum achievable Purcell factor for each single mode

$$F_p^{max} = \frac{3}{4\pi^2} \left(\frac{\lambda}{n}\right)^3 \frac{Q}{V} \approx 19 \tag{2}$$

that can only be achieved for a monochromatic emitter, located at the antinode of the mode, in perfect spectral resonance with the mode and with a dipole aligned with the local polarization of the field at the antinode (i.e. an in plane dipole for these CBG cavities). [19]

We consider a single color center with a dipole vector $\boldsymbol{d}$ oriented along $\langle 111 \rangle$, located at position $\boldsymbol{r}$ in the resonant electric field $\boldsymbol{E}$. The resonant mode is spectrally tuned with the ZPL : $\delta = \lambda_{res} - \lambda_{ZPL} = 0$. The Purcell factor experienced by the color center reads

$$F_p(\boldsymbol{r}) = F_p^{max} \times \sum_{i=1,2} \frac{|\boldsymbol{d}.\boldsymbol{E}_i(\boldsymbol{r})|^2}{|\boldsymbol{d}|^2 max(|\boldsymbol{E}_i|^2)} \tag{3}$$

where $i$ is an index for the two polarized modes. The zero-phonon emission rate into the resonant mode is equal to $\eta_{QE} \xi F_p(\boldsymbol{r})/\tau_{bulk}$, where $\eta_{QE}$ is the quantum yield of the emitter, $\xi \approx 39\%$ is the Debye-Waller factor and $\tau_{bulk}$ is the lifetime of the emitter in bulk silicon. The emission rate into the continuum of other modes (« leaky modes ») is modified by a factor $\gamma$ compared to the emission rate in bulk silicon, and thus reads $\eta_{QE}\gamma/\tau_{bulk}$. The non-radiative recombination rate $(1 - \eta_{QE})/\tau_{bulk}$ is assumed to be the same in the cavity and bulk silicon. The total decay rate in the cavity $1/\tau_{cav}(\boldsymbol{r})$ normalized to the decay rate in the bulk material $1/\tau_{bulk}$ is thus equal to

$$\frac{1/\tau_{cav}(\boldsymbol{r})}{1/\tau_{bulk}} = \eta_{QE}\xi F_p(\boldsymbol{r}) + \eta_{QE}\gamma + (1 - \eta_{QE}) \tag{4}$$

After a weak excitation pulse, the population of the excited state decays in the available recombination channels. Importantly, the total number of recombination events following the pulse does not depend upon detuning as it is only defined by the weak excitation probability of the emitters. Nevertheless, Purcell enhancement of spontaneous emission increases the proportion of radiative recombinations, resulting in higher emitted intensity. [45] The proportion of zero-phonon recombinations includes radiative emission both in the resonant and leaky modes. For the first channel, the photons are collected with an efficiency $\eta_{mode} = 40\%$, while for the second one the efficiency is much smaller $\eta_{leaks} = 5 \pm 2$ %. Under these considerations, the intensity of the ZPL for a single W center at position $\boldsymbol{r}$ reads:

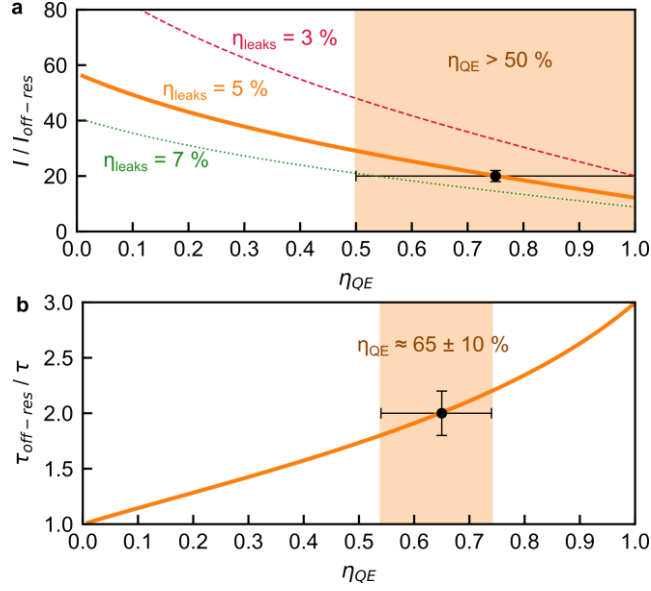

**Figure 4.** Modeling of Purcell enhancement for an ensemble of W centers. **a.** Intensity enhancement ratio and **b.** PL decay acceleration ratio as a function of the quantum yield $\eta_{QE}$ of W centers in bulk silicon.

$$I_{1W}^{ZPL}(\mathbf{r}) \propto \frac{\eta_{mode}\eta_{QE}\xi F_p(\mathbf{r}) + \eta_{leaks}\eta_{QE}\xi\gamma}{\eta_{QE}\xi F_p(\mathbf{r}) + \eta_{QE}\gamma + (1 - \eta_{QE})} \quad (5)$$

The total collected intensity is the sum of the intensity collected from each emitter,

$$I_{tot}^{ZPL} = \int \rho(\mathbf{r}) I_{1W}^{ZPL}(\mathbf{r}) d\mathbf{r} \quad (6)$$

where $\rho(\mathbf{r})$ is the spatial distribution of the emitters in the cavity. From PL raster scans, $\rho(\mathbf{r})$ is considered uniform in the plane of the silicon layer. Using an experimental technique of digital etching of the Si layer, the distribution of the W centers along the depth of the sample was estimated to a Gaussian profile $\propto \exp(-0.5\,(z - z_0)^2/\sigma^2)$ with $z_0 = 127 \pm 6$ nm and $\sigma = 35 \pm 5$ nm, where $z$ is the distance from the Si/SiO$_2$ interface (see the **Supplementary Information**).

**Figure 4a** displays the calculated ratio between the ZPL intensity on-resonance and off-resonance as a function of the reference quantum efficiency of W in bulk Si, for various values of $\eta_{leaks}$ taken in the 3 – 7 % range. The Purcell-enhanced proportion of zero-phonon emission compared to the other recombination channels results in an increase of the ZPL intensity. This exaltation is larger when the proportion of zero-phonon emission is initially low, which is the case when the quantum efficiency is poor. The experimental factor of 20, represented as a black dot in the figure, is consistent with our model for $\eta_{QE} > 50\%$. Note that this estimation is based solely on PL intensity and thus strongly depends on $\eta_{leaks}$ that is known with limited precision. Nevertheless, these results show that our model is able to predict the order of magnitude of the cavity-enhanced PL intensity.

The time-resolved intensity collected after the excitation pulse is modelled as the sum of the exponential decay of each emitter, weighted by the intensity collected from each emitter:

$$I_{tot}^{coll}(t) = \int \rho(\mathbf{r}) I_{1W}^{coll}(\mathbf{r}) \exp\left(-\frac{t}{\tau_{cav}(\mathbf{r})}\right) d\mathbf{r} \quad (7)$$

where $I_{1W}^{coll}(\boldsymbol{r}, \delta)$ is the intensity collected in the 1200 – 1300 nm spectral range. The calculated decay curves are shown in **Figure S5** in the **Supplementary Information**. From Equation (7), one can define the average lifetime as a function of detuning as

$$\langle \tau \rangle = \frac{\int \rho(\boldsymbol{r}) I_{1W}^{coll}(\boldsymbol{r}) \tau_{cav}(\boldsymbol{r}) d\boldsymbol{r}}{\int \rho(\boldsymbol{r}) I_{1W}^{coll}(\boldsymbol{r}) d\boldsymbol{r}} \qquad (8)$$

The ratio between the decay rate on-resonance and off-resonance is shown in **Figure 4b**. The Purcell-enhancement of zero-phonon emission into the resonant mode results in an increase of the total decay rate. This acceleration is all the stronger as zero-phonon emission initially dominates the recombination dynamics, i.e. when the quantum efficiency is high. The experimental factor of $2 \pm 0.2$ is consistent with our model for $\eta_{QE} = 65 \pm 10~\%$, which is compatible with the intensity-based analysis. This value compares favorably to previously reported values for the G center. [28,29]

**CONCLUSIONS**

In this work, we perform CQED experiments in an all-silicon atom-resonator system consisting of W centers in CBG cavities. We demonstrate Purcell enhancement of the collected ZPL intensity by a factor of 20, together with an acceleration of the average PL decay rate by a factor of 2. Although the emitters exhibit an uncommon power-dependent decay rate, the exaltation factors are weakly dependent over power. We present a theoretical model of Purcell enhancement for an ensemble of W centers and estimate the quantum efficiency of the present ensemble to be $65 \pm 10~\%$.

Our modeling highlights the crucial role of quantum efficiency in CQED experiments in a quantitative way. This issue generally does not need consideration, [19,47] as quantum emitters such as semiconductor QDs at cryogenic temperatures and NV centers in diamond have a quantum efficiency close to unity. In the case of W centers, it will be worth exploring other fabrication methods as they may allow to create W centers with higher radiative quantum yield. For example, recent reports about the G centers tend to show that their optical properties strongly depend upon their creation process. [29,48]

Nevertheless, our results are already very encouraging in view of CQED experiments with a single W center maximally coupled to a CBG cavity, as highlighted in **Table 1**. By selectively enhancing zero-phonon emission and additionally inhibiting phonon-assisted emission, the cavity erases to a large extend the imperfections of the W center, whose quantum efficiency and Debye-Waller factors reach respectively 0.9 and 0.97. Additionally, the cavity induces a significant lifetime shortening (from 34 ns to 9.2 ns). Considering finally spontaneous emission control, we see that ZPL photons are very efficiently funneled into the cavity mode as $\beta = F_p / (F_p + \gamma) \approx 0.98$.

**Table 1.** Relevant recombination rates and figures of merit for a W center in bulk Si, or optimally-coupled to the resonant mode of a CBG cavity ($F_p = 12.5$, $\gamma = 0.27$).

|  | Zero-phonon emission rate | Phonon-assisted emission rate | Non-radiative recombination rate | Lifetime | Quantum efficiency | Debye-Waller factor |
|---|---|---|---|---|---|---|
| W in bulk Si | 1 / 134 ns$^{-1}$ | 1 / 86 ns$^{-1}$ | 1 / 97 ns$^{-1}$ | 34 ns | 0.65 | 0.39 |
| W in CBG cavity | 1 / 10.5 ns$^{-1}$ | 1 / 318 ns$^{-1}$ | 1 / 97 ns$^{-1}$ | 9.2 ns | 0.90 | 0.97 |

Since single W centers have already demonstrated highly pure single photon emission and perfect linear polarization of emitted photons, these figures of merit show that single W centers in CBG cavities are already interesting for building efficient, off-chip emitting sources of polarized single photons. They are also well adapted to a combination with high-$F_p$ photonic crystal cavities coupled to single-mode waveguides, so as to build efficient SPS for quantum photonic circuits. The experimental realization of such SPS will require positioning a single W center precisely at the mode antinode of the CBG or photonic crystal cavity. [49] In this context, localized creation of W centers using focused ion beam implantation has already been demonstrated. [33]

**METHODS**

The sample was held in a cryostat and the measurements were performed using a confocal microscopy setup. Pulsed and CW optical excitation were provided by a 485 nm laser diode focused onto the sample using a microscope objective of numerical aperture $NA = 0.26$ (Mitutoyo M Plan Apo NIR 10X). The temporal width of the pulses is of ~100 ps. The PL was collected using the same objective, and achieved to a fiber-coupled superconducting nanowire single photon detector (SNSPD), or to a grating spectrometer with a photodiode array camera for spectral analysis. For measurement of the ZPL intensity as a function of detuning, the spectrometer was used to select a 0.5 nm spectral window around the ZPL and the output PL was sent to the SNSPD. For time-resolved measurements, the PL was spectrally filtered with a 1200 nm longpass filter and a 1300 nm shortpass filter. For PL raster scans, the excitation was provided by a CW laser diode emitting at 532 nm focused onto the sample using an objective of numerical aperture $NA = 0.85$. The excitation spot was swept across the sample using a steering mirror. The spatial resolution of the scans is around 1 μm.

**ACKNOWLEDGEMENTS**

This work is supported by the French National Research Agency (ANR) through the "OCTOPUS project n° ANR-18-CE47-0013" and by the PTC-MP "W-TeQ" internal project of CEA. B. Lefaucher is supported by the "Program QuantForm-UGA n° ANR-21-CMAQ-003 France 2030" and "Laboratoire d'Excellence LANEF n° ANR-10-LABX-51-01". The authors acknowledge the assistance of the staff of the Grenoble Advanced Technological Platform (PTA).

# Purcell enhancement of silicon W centers in circular Bragg grating cavities

## Supplementary Information

Baptiste Lefaucher[1]*, Jean-Baptiste Jager[1], Vincent Calvo[1], Félix Cache[2], Alrik Durand[2], Vincent Jacques[2], Isabelle Robert-Philip[2], Guillaume Cassabois[2], Yoann Baron[3], Frédéric Mazen[3], Sébastien Kerdilès[3], Shay Reboh[3], Anaïs Dréau[2] and Jean-Michel Gérard[1]*

[1]Université Grenoble Alpes, CEA, Grenoble INP, IRIG, PHELIQS, Grenoble, France

[2]Laboratoire Charles Coulomb, Université de Montpellier and CNRS, Montpellier, France

[3]Université Grenoble Alpes, CEA-LETI, Minatec Campus, Grenoble, France

* baptiste.lefaucher@cea.fr, jean-michel.gerard@cea.fr

**Far-field radiation pattern and collection efficiency**

The far-field radiation pattern of the ZPL-resonant CBG cavity is shown in **Figure S1a**. The collection efficiency $\eta_{coll}$, defined as the collected power normalized to the total emitted power, is plotted as a function of numerical aperture in **Figure S1b**. The figure also shows the fraction of upwards emitted power that is collected, which is equal to $\eta_{coll}/\eta_{z+}$ where $\eta_{z+}$ is the upwards extraction efficiency.

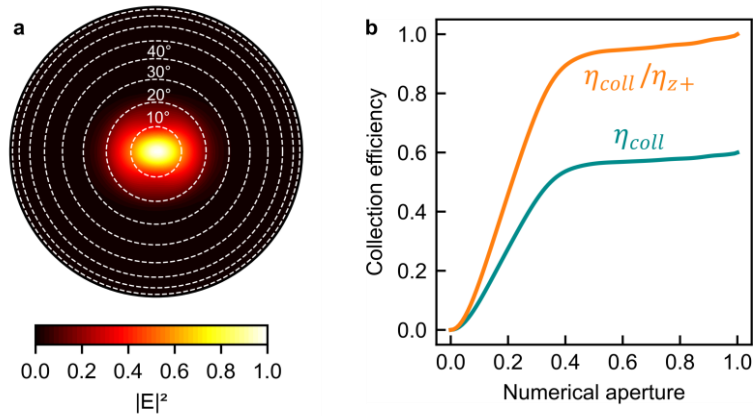

**Figure S1.** Far-field radiation pattern and collection efficiency of the ZPL-resonant cavity. **a.** Far-field radiation pattern. **b.** Collection efficiency $\eta_{coll}$ and fraction of collected upwards emission $\eta_{coll}/\eta_{z+}$ as a function of numerical aperture.

**TRPL with spectral filtering**

**Figure S2** shows the PL decay curves that were obtained for the reference 2D layer under different filtering conditions. The decay curves are nearly identical for all filtering conditions, showing that the slow temporal components cannot be attributed to parasitic emitters and must be considered as properties of the present ensemble of W centers.

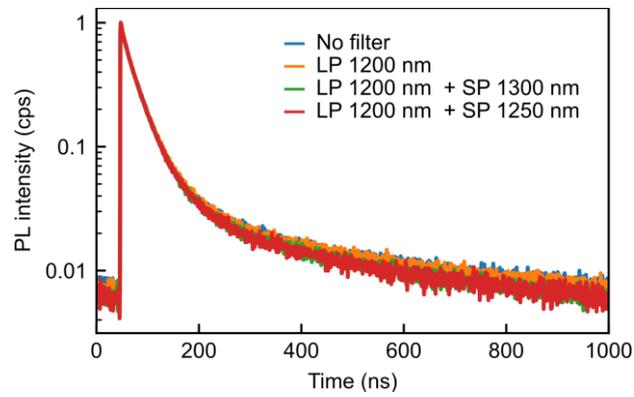

**Figure S2.** PL decay curves obtained under different spectral filtering conditions for the reference 2D layer. LP : longpass ; SP : shortpass.

**Power study**

**Figure S3a** shows the power characteristics of the emitters under pulsed excitation for a ZPL-resonant and off-resonant cavity, and for the 2D layer of reference. The data fit fairly well to the saturation model for two-level emitters, $I = I_{sat}/(1 + (P_{sat}/P))$ where $P_{sat}$ and $I_{sat}$ are the saturation power and intensity at saturation for a single emitter, respectively. The straight lines in the figure correspond to the linear power regime of the emitters, showing that the W centers are far from saturation at the working power of 0.95 µW.

PL intensity decay curves obtained for different power values in the 2D layer are shown in **Figure S3b**. As power increases, a short time component arises and the average recombination rate increases, which is unexpected for two-level quantum emitters. **Figure S3c** shows the power-dependent recombination rate for the two cavities and 2D layer. For the cavities, the decay rate increases with power up to 5 µW, then slightly decreases. For the 2D layer, the decay rate increases with power and tends to reach a saturation value. The data fit well to a function of the form $1/\tau = 1/\tau_0 + 1/\tau_1 \, [P/(P + P_1)]$, with $1/\tau_0 = 0.027 \pm 0.001 \text{ ns}^{-1}$, $1/\tau_1 = 0.020 \pm 0.001 \text{ ns}^{-1}$ and $P_1 = 7.1 \pm 0.5$ µW. This behavior is tentatively attributed to power-induced thermal heating. A thorough investigation of this phenomena is beyond the scope of this paper; nevertheless, an important point is that the laser pump power might affect the outcome of the time-resolved CQED experiments. The acceleration factor $\tau_{off-res}/\tau_{ZPL-res}$ is plotted as a function of power in **Figure S3d**. This factor reaches ~1.85 for powers below 3 µW and decreases by 10 % for higher power values. We can conclude that the PL decay acceleration due to the coupling with the cavity mode is clearly observed over the full power range under study, with an acceleration factor almost power-independent.

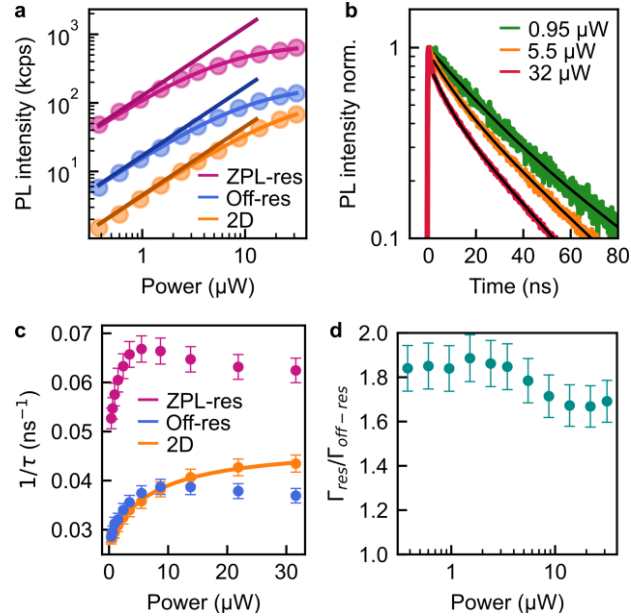

**Figure S3.** Power dependency of the recombination rate. **a.** Power characteristics of the ZPL-resonant cavity, off-resonant cavity and 2D layer under 1 MHz pulsed excitation. **b.** Decay curve of the 2D layer for different power values. **c.** Recombination rate and **d.** Acceleration ratio $\Gamma_{res}/\Gamma_{off-res}$ as a function of excitation power.

## Spatial distribution of the W centers

**Figure S4a** presents a PL raster scan of the 2D layer, showing bright emission all over a $15 \times 15 \; \mu m^2$ area. We could thus assume in the modelisation that the ensemble of W centers is dense and uniformly distributed along the surface of the sample.

The distribution of the W centers along the depth in the layer was estimated in a digital etching experiment. A sample was cut into six pieces, five of which were etched in order to obtain different Si layer thicknesses. The six samples had a Si layer thickness of 38, 65, 100, 135, 165 and 220 nm, respectively as determined after inspection by scanning electron microscopy. **Figure S4b** shows the normalized PL intensity as a function of layer thickness. The data fit reasonably well to a cumulative distribution function of Gaussian distribution

$$I(d) = \frac{1}{2}\left[1 + \text{erf}\left(\frac{d - d_0}{\sigma\sqrt{2}}\right)\right] \quad (S1)$$

where erf is the error function, with $d_0 = 127 \pm 6$ nm and $\sigma = 35 \pm 5$ nm. Assuming the PL intensity is only dependent upon the number of emitters in the layer, the W centers are distributed according to a Gaussian profile of average altitude $d_0$ (distance from the Si/SiO$_2$ interface) and standard deviation $\sigma$, as shown in **Figure S4c**. The grey curve corresponds to the distribution of implanted silicon interstitial atoms calculated by SRIM. These results show that the W centers are mostly located close to the targeted depth of the implanted ions, at the center of the Si layer.

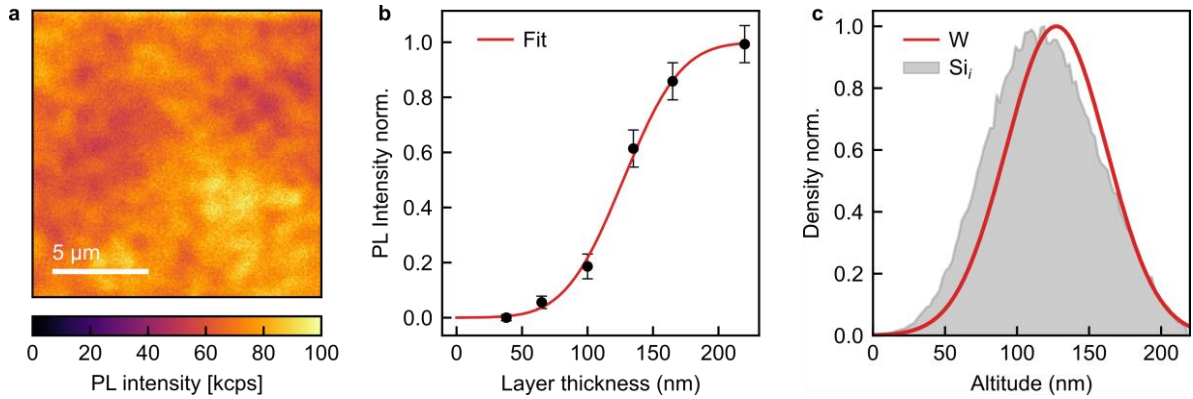

**Figure S4.** Spatial distribution of the emitters. **a.** PL raster scan of the 2D layer. **b.** PL intensity as a function of the silicon layer thickness. **c.** Distribution of the W centers estimated from the digital etching experiment (red), compared to the distribution of implanted silicon interstitial atoms calculated by SRIM (grey).
19

**Calculated decay of an ensemble of W centers in a CBG cavity**

**Figure S5a** shows the distribution of lifetimes of the emitters in the cavity for a given off-resonant time $\tau_{off-res}$, for several quantum efficiency values. The resulting decay curves, calculated using Equation (7), are displayed in **Figure S5b**, showing a multi-exponential behaviour due to the high dispersion of lifetimes.

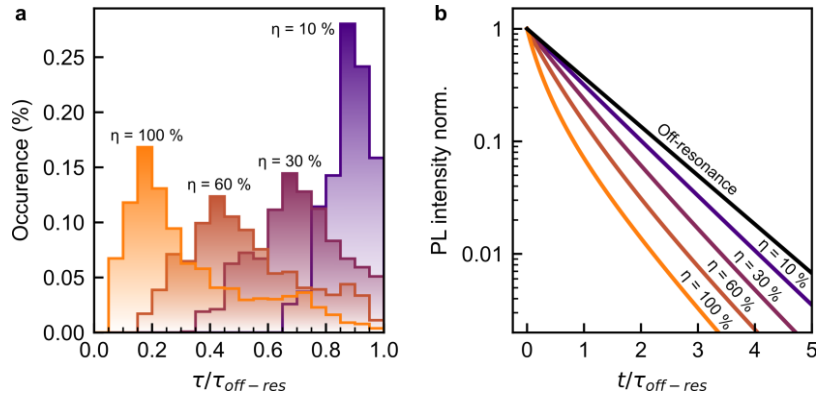

**Figure S5.** PL decay of an ensemble of W centers in a CBG cavity. **a.** Distribution of the lifetimes of the emitters. **b.** Calculated PL decay curves.